\documentclass[twocolumn]{aa} 
\usepackage{graphicx}
\usepackage{multirow}
\usepackage{txfonts}
\usepackage{amsmath} 
\newcommand{\logxion}{log($\xi_{ion}^*$/Hz erg$^{-1}$)}
\newcommand{\xion}{$\xi_{ion}^*$}
\newcommand{\ciii}{CIII]$\lambda 1909$}

\newcommand{\heii}{HeII$\lambda 1640$}
\newcommand{\oiii}{OIII]$\lambda 1666$}
  
\begin{document} 

\title{The ionizing photon production efficiency of bright z$\sim$2-5 galaxies}
   \author{M. Castellano\inst{1}, D. Belfiori\inst{1}, L. Pentericci\inst{1}, A. Calabr\`{o}\inst{1}, S. Mascia\inst{1}, L. Napolitano\inst{1}, F. Caro\inst{1}, S. Charlot\inst{2}, J. Chevallard\inst{3}, E. Curtis Lake\inst{4},  M. Talia\inst{5,6}, A. Bongiorno\inst{1}, A. Fontana\inst{1}, J. P. U. Fynbo\inst{7,8}, B. Garilli\inst{9}, L. Guaita\inst{10}, R. J. McLure\inst{11}, E. Merlin\inst{1},  M. Mignoli\inst{6}, M. Moresco\inst{6}, E. Pompei\inst{12}, L. Pozzetti\inst{6}, A. Saldana Lopez\inst{13}, A. Saxena\inst{14,15}, P. Santini\inst{1}, D. Schaerer\inst{13}, C. Schreiber\inst{16}, A. E. Shapley\inst{17}, E. Vanzella\inst{6}, G. Zamorani\inst{6}}
   \institute{
          $^{1}$INAF - OAR, via Frascati 33, 00078 Monte Porzio Catone (Roma) - Italy\\
          $^{2}$Sorbonne Universit\'e, CNRS, UMR 7095, Institut d’Astrophysique de Paris, 98 bis bd Arago, 75014 Paris, France\\
          $^{3}$Department of Physics, University of Oxford, Denys Wilkinson Building, Keble Road, Oxford OX1 3RH, UK\\
          $^{4}$Centre for Astrophysics Research, Department of Physics, Astronomy and Mathematics, University of Hertfordshire, Hatfield, AL10 9AB, UK\\
          $^{5}$University of Bologna - Department of Physics and Astronomy “Augusto Righi” (DIFA), Via Gobetti 93/2, I-40129 Bologna, Italy\\
          $^{6}$INAF -- OAS, Osservatorio di Astrofisica e Scienza dello Spazio di Bologna, via Gobetti 93/3, I-40129 Bologna, Italy\\
          $^{7}$Cosmic Dawn Center (DAWN), Niels Bohr Institute, University of Copenhagen, Jagtvej 128, 2200 Copenhagen, Denmark\\
          $^{8}$Niels Bohr Institute, University of Copenhagen, Blegdamsvej 17, DK2100 Copenhagen Ø, Denmark
          $^{9}$INAF - Istituto di Astrofisica Spaziale e Fisica Cosmica, Via A. Corti 12, 20133 Milan, Italy\\
          $^{10}$Instituto de Astrofisica, Facultad de Ciencias Exactas, Universidad Andres Bello, Fernandez Concha 700, Las Condes, Santiago RM, Chile\\
          $^{11}$Institute for Astronomy, University of Edinburgh, Royal Observatory, Edinburgh EH9 3HJ, UK\\
          $^{12}$European Southern Observatory, Alonso de Córdova 3107, Vitacura, Santiago de Chile, Chile
          $^{13}$Department of Astronomy, University of Geneva, 51 Chemin Pegasi, 1290 Versoix, Switzerland\\
          $^{14}$Sub-department of Astrophysics, University of Oxford, Keble Road, Oxford OX1 3RH, UK\\
          $^{15}$Department of Physics and Astronomy, University College London, Gower Street, London WC1E 6BT, UK\\
          $^{16}$IBEX Innovations, Sedgefield, Stockton-on-Tees, TS21 3FF, United Kingdom\\
          $^{17}$Department of Physics \& Astronomy, University of California, Los Angeles, 430 Portola Plaza, Los Angeles, CA 90095, USA
             }

   \date{...}

 
  \abstract
   {}
   {We investigate the production efficiency of ionizing photons (\xion) of 1174 galaxies with secure redshift at z=2-5 from the VANDELS survey to determine the relation between ionizing emission and physical properties of bright and massive sources.}
   {We constrain \xion~and galaxy physical parameters by means of spectro-photometric fits performed with the \texttt{BEAGLE} code. The analysis exploits the multi-band photometry in the VANDELS fields, and the measurement of UV rest-frame emission lines (\ciii, \heii, \oiii) from deep VIMOS spectra.}
   {We find no clear evolution of \xion~with redshift within the probed range.  The ionizing efficiency slightly increases at fainter $M_{UV}$, and bluer UV slopes, but these trends are less evident when restricting the analysis to a complete subsample at log(M$_{star}$/M$_{\odot}$)$>$9.5. We find a significant trend of increasing \xion~with increasing EW(Ly$\alpha$), with an average \logxion$>$25 at EW$>$50\AA, and a higher ionizing efficiency for high-EW \ciii~and \oiii~emitters. 
   The most significant correlations are found with respect to stellar mass, specific star-formation rate (sSFR) and SFR surface density ($\Sigma_{SFR}$). The relation between \xion~and sSFR shows a monotonic increase from \logxion  $\sim$24.5 at log(sSFR)$\sim$-9.5$yr^{-1}$ to $\sim$25.5 at log(sSFR)$\sim$-7.5$yr^{-1}$, a low scatter and little dependence on mass. The objects above the main-sequence of star-formation consistently have higher-than-average \xion. A clear increase of \xion~with $\Sigma_{SFR}$ is also found, with \logxion$>$25 for objects at $\Sigma_{SFR}>$10 M$_{\odot}/yr/kpc^2$.}
   {Bright ($M_{UV}\lesssim$20) and massive (log(M$_{star}$/M$_{\odot}$)$\gtrsim$9.5) galaxies at z=2-5 have moderate ionizing efficiency. However, the correlation between \xion~and sSFR, together with the known increase of average sSFR with redshift at fixed stellar mass, suggests that similar galaxies in the epoch of reionization can be efficient sources of ionizing photons. The availability of sSFR and $\Sigma_{SFR}$ as proxies for \xion~can be of fundamental importance to determine the role at the onset of reionization of galaxy populations at z$\gtrsim$10 recently discovered by JWST.}
   
\authorrunning{M. Castellano et al.}   
\titlerunning{The ionizing efficiency of bright z$\sim$2-5 galaxies}   

\maketitle

\section{Introduction}

Investigating the ionizing emission of star-forming galaxies is key to understanding how galaxies form stars and affect the surrounding environment. In particular, it is fundamental to constrain the role played by high-redshift galaxies in the epoch of reionization of the inter-galactic medium at z$\gtrsim$6 \citep[EoR,][]{Dayal2018,Robertson2022}. 
The rate of ionizing photons escaping into the IGM from a galaxy population is $\dot{N}=\rho_{UV}\xi_{ion}^*f_{esc}$, where $\rho_{UV}$ is the UV luminosity density,  $\xi_{ion}^*$ is the ionizing photon production efficiency per unit UV luminosity, and $f_{esc}$ is the fraction of ionizing photons leaked into the galaxy surroundings.  The $\rho_{UV}$ can be constrained by measuring the galaxy UV luminosity function \citep[e.g.,][]{Castellano2010b,Bouwens2021}. The escape fraction of ionizing photons can only be directly constrained at $z\lesssim$3-4 \citep[e.g.,][]{Vanzella2016,Marchi2017,Steidel2018,Pahl2021}, but the recent findings on indirect estimators of $f_{esc}$ in low-redshift galaxies \citep[e.g.,][]{Izotov2018a,Izotov2018b,Flury2022} is enabling the first constraints on the escape fraction of galaxies in the EoR \citep[e.g.,][]{Lin2023,Mascia2023b}.

To constrain the contribution of high-redshift galaxies to the EoR, much effort has also been spent to study the ionizing photon production efficiency $\xi_{ion}^*$ as a function of redshift and of galaxy properties. 

Theoretical models predict a mild evolution of $\xi_{ion}^*$ with redshift, with typical values in the reionization epoch in the range \logxion$\simeq$25.1-25.5. The ionizing photon production efficiency is predicted to be a factor of $\sim$2 higher in low-mass galaxies due to their lower metallicity, which likely drive trends of increasing \xion with decreasing SFR, UV slope and UV luminosity and increasing sSFR \citep[]{Wilkins2016,Ceverino2019,Yung2020}. However, model predictions have been found to critically depend on the adopted stellar population synthesis (SPS) models, with the inclusion of binary stellar populations increasing \xion by a factor of $\gtrsim$2 in simulated galaxies \citep[e.g,][]{Ma2016,Yung2020}.
 
Direct observational constraints on the ionizing efficiency \xion~can be obtained from the spectroscopic measurement of Balmer emission lines after correcting for dust attenuation \citep[e.g.,][]{Schaerer2016,Shivaei2018}. Similarly, photometric measurements of the flux from optical emission lines can be used when spectroscopic observations are not available \citep[e.g.,][]{Bouwens2016a}. As an alternative, it has been shown that \xion~can be estimated from the equivalent width (EW) of the [OIII]$\lambda 4959,5007$ doublet \citep[][]{Chevallard2018,Reddy2018,Tang2019}, rest-frame UV colors \citep[][]{Duncan2015} or UV rest-frame emission lines \citep[][]{Stark2015a}. 
The reference ionizing efficiency assumed in reionization scenarios is log($\xi_{ion}^*$/Hz erg$^{-1}$)$\simeq$25.2-25.3 \citep[][]{Robertson2013}, which is consistent with the value measured in Lyman-break galaxy (LBG) samples at z$\sim$4-5 and with predictions from theoretical models \citep[][]{Wilkins2016}.

However, measurements show a wide range of values for different classes of objects, ranging from the $\sim$24.8 of local compact star-forming galaxies \citep[CSFGs][]{Izotov2017} and z$\sim$2 H-$\alpha$ emitters \citep[][]{Matthee2017}, to extreme log($\xi_{ion}^*$/Hz erg$^{-1}$)$\simeq$26.3 of faint Lyman-$\alpha$ emitters at z$\sim$4-5. Higher-than-average ionizing efficiencies log($\xi_{ion}^*$/Hz erg$^{-1}$)$\gtrsim$25.5 have also been found in local Lyman-continuum leakers \citep[][]{Schaerer2016}, Lyman-$\alpha$ emitters at z$\sim$3-5 \citep[][]{Harikane2018,Sobral2019}, strong line emitters at z$>$2 \citep[][]{Nakajima2016,Tang2019}. It has been shown that \xion~remains approximately constant as a function of observed UV luminosity at fixed redshift, while it increases in objects with blue UV slopes \citep[][]{Bouwens2016a,Shivaei2018,Lam2019,Izotov2021}. A correlation between \xion~and specific star-formation-rate (sSFR) has been found by \citet[][]{Izotov2021} on CSFGs at z$\leq$1, and is apparent at higher redshifts from the correlation between \xion~and EW($H\alpha$) found by \citet[][]{Faisst2019} (3.9$<z<$4.9), \citet[][]{Emami2020} (1.4$<z<$2.7),  and \citet[][]{Prieto-Lyon+22} (3$<z<$7). 

Few constraints are available for galaxies in the EoR. Some objects at $z\gtrsim7$ have been found to have very high ionizing efficiencies \citep[$\gtrsim$25.7, e.g.,][]{Stark2015b,Stark2017,Endsley2021b,Endsley2022c,Stefanon2022,Fujimoto2023}. In other cases, the \xion has been estimated to be consistent or slightly higher than the canonical range assumed for high-redshift galaxies \citep[][]{Castellano2022a,Schaerer2022}.
The suggested trend of an increasing typical \xion~with redshift likely arises from a change in the underlying mixture of galaxy populations, eventually driven by an evolution in physical parameters. Similarly, discrepancies among \xion~estimates can result from the different sampling of the various galaxy populations by the different selection techniques. In fact, \xion~has been found to increase in objects with high ionization state, young ages and low metallicity, and is also affected by star-formation burstiness, initial-mass function (IMF) and evolution of stellar populations \citep[][]{Shivaei2018,Chisholm2019}.
It is thus fundamental to investigate the relation between galaxy properties and ionizing efficiency in order to fully constrain the role of star-forming galaxies in the EoR.

In the present paper we exploit a large sample of galaxies at z$\sim$2-5 from the VANDELS survey \citep[][]{McLure2018,Pentericci2018} provided with robust measurements of their spectroscopic redshift, spectral energy distribution (SED) and UV emission lines, to investigate the relation between \xion~and galaxy physical parameters.   
The paper is organised as follows: the sample is presented in Sect.~\ref{sec:sample}, while in Sect.~\ref{sec:BEAGLE} we discuss the spectro-photometric fit used to constrain \xion~and other physical properties. The analysis of how \xion~varies with observed properties and physical parameters is presented in Sect.~\ref{sec:results}. 
The results are summarised in Sect.~\ref{sec:summary}. Throughout the paper we adopt AB magnitudes \citep{Oke1983}, a \citet[][]{Chabrier2003} IMF and a $\Lambda$-CDM concordance model ($H_0$ = 70 km s$^{-1}$ Mpc$^{-1}$, $\Omega_M=0.3$, and $\Omega_{\Lambda}=0.7$).

\section{The VANDELS sample}\label{sec:sample}

For our work we use data  from VANDELS, a recently completed ESO public spectroscopic survey carried out using the VIMOS spectrograph on the Very Large Telescope (VLT). VANDELS targets were selected in the \textit{UKIDSS Ultra Deep Survey} (UDS), and the \textit{Chandra Deep Field South} (CDFS). VANDELS footprints are centered on the  HST areas observed by  the  CANDELS program \citep{Grogin2011,Koekemoer2011}, but  given the  VIMOS field of view, the areas observed are two times larger: for the outer areas, not covered by the HST imaging data, new photometric catalogs were assembled.

The survey description and initial target selection strategies are described  in \citet{McLure2018}, while data reduction and redshift determination  can be found in \cite{Pentericci2018}. The fourth and final data release that we use is fully described in \cite{Garilli2021} and contains redshifts of $\sim$2100 sources and the above mentioned photometric catalogs\footnote{The data release is available at \url{http://vandels.inaf.it/db} and \url{https://www.eso.org/qi/}}. In particular the  redshifts were derived using the \texttt{pandora.ez} tool \citep{Garilli2010} which cross-correlates the spectra with a series of  galaxy templates, and then checked visually by 4 different members of the collaboration. The reliability  of the redshifts is quantified with a quality flag (QF) in the following way: 0 means no redshift could be determined; 1 indicates a  50\% probability of the redshift being correct; 2 indicates a 70/80\% probability of the redshift being correct; 3 and 4 indicate respectively  a 95 and 100\% probability of the redshift being correct; finally  9 means that the spectrum shows a single emission line and that the redshift corresponds to the most probable identification. In this work we have analysed all sources with $2\leq z \leq 5$ and secure redshift, i.e. $QF = 3$ and $=4$.

We exclude 13 AGN selected as in Bongiorno et al. (in prep). The objects  have been considered as likely AGN because they either had an X-ray counterpart from \cite{Luo2017} or \cite{Kocevski2018} (for the CDFS and UDS fields respectively) within a  1.5$''$ radius, showed typical broad emission lines, or showed high ionization narrow emission lines with line ratios typical of AGN   \citep[according to the diagnostics described in][]{Feltre2016}.  
The final sample comprises 1174 galaxies, of which 604 are in CDFS and 577 are in the UDS field. All objects were selected as star-forming galaxies or LBGs in the VANDELS target preparation, except three sources initially classified as potential AGN on the basis of their SED, and one Herschel-detected source. We verified that their inclusion is not affecting the analysis presented here.

\subsection{Photometric measurements}\label{subsec:photometry}

For the objects in our sample we use the VANDELS photometric catalogs described by \citet[][]{McLure2018} for the outer CDFS and UDS areas, the CANDELS photometric catalog by \citet[][]{Galametz2013} for the UDS-HST field, and the improved CANDELS catalog including both photometry from \citet[][]{Guo2013} and the Ks-band HAWKI data from \citet[][]{Fontana2014} for the CDFS-HST one. The objects in our sample are provided with a measurement of their UV slope performed by fitting the available broad-band photometric measurements sampling the 1230–2750\AA~rest-frame range as described in \citet[][]{Calabro2021}. We find that the UV slope is robustly measured when at least 3 photometric bands are used in the fit, and when the uncertainty on the measurement is $\sigma(\beta)<$0.5. This condition is met for 810 objects out of the 1174 in the parent sample. For each of these objects the UV slope determination is also used to estimate the rest-frame UV magnitude, $M_{UV}$, from a simple interpolation of the continuum at 1600\AA. Finally, we use the equivalent radii measured by \citet[][]{Calabro2022a} for 749 objects covered by F814W HST imaging. This subsample has a median F814W magnitude of 25.2, corresponding to a signal-to-noise ratio (SNR) $\sim$20, enabling an accurate measurement of their size \citep[see also][for the relevant methodology]{Ribeiro2016}.

\subsection{Spectroscopic measurements}\label{subsec:spectra}
The measurements we use come from the official VANDELS spectroscopic catalog (Talia et al. in prep.). Specifically, we use a Gaussian fit obtained by \texttt{slinefit} \citep[][]{Schreiber2018} for the Ly$\alpha$ line and direct integration measurements for the other UV emission lines. We visually inspected the spectra in the Ly$\alpha$ range and assigned a flag to indicate whether a single Gaussian provides a good fit to the line or not. In the present paper we will use EW(Ly$\alpha$) only for the 565 objects in our sample with a robust Gaussian fit of the line.

The direct integration measurements have been performed using \texttt{PyLick}, a tool developed to measure Lick-like indices and continuum breaks extensively tested on galaxies from the LEGA-C survey \citep[][]{vanderWel2016} and on VANDELS sources  \citep[][]{Borghi2022}. The parameters to define are the integration windows and the bandpasses for the blue and red part of the continuum.
While for absorption lines the bandpasses were already defined in the literature \citep[][]{Maraston2009,Leitherer2011}, for emission lines they are not standard and were newly defined on the basis of a high SNR ($\sim$35/pixel) composite spectrum of all sources with  VANDELS quality flag 3 and 4 (see Talia et al. for more details).

In the present paper we will use measurements of \heii, \oiii~and \ciii. We did not consider CIV$\lambda 1548$ because it is a blend of stellar and nebular emission with a profile resulting from the mixture of both emission and absorption features. The properties of VANDELS CIV-emitters, including their ionizing efficiency, are discussed in a companion paper \citep[][]{Mascia2023}.
For the UV lines of interest the following windows were used:
$1634-1654$\AA~for the \heii~emission  
(blue and red continuum windows at $1614-1632$, and $1680-1705$\AA~respectively); $1663-1668$\AA~for  \oiii~ (continuum: $1614-1632$, and $1680-1705$\AA); $1897-1919$\AA~for \ciii~ (continuum: $1815-1839$, and $1932-1948$\AA~respectively).  The uncertainties on the measurements are evaluated by \texttt{PyLick} following the SNR method by \citet[][]{Cardiel1998}. Talia et al. found that the error spectra produced by the data reduction pipeline underestimate the noise level by a factor of $\sim$2, on average, with respect to the noise r.m.s. measured on the object spectra. Therefore, they applied an a-posteriori correction factor to the error spectra that are used to derive measurement uncertainties.

\begin{figure}
\centering
\includegraphics[trim={1.5cm 1.cm 2.cm 2.cm},clip,width=\linewidth,keepaspectratio]{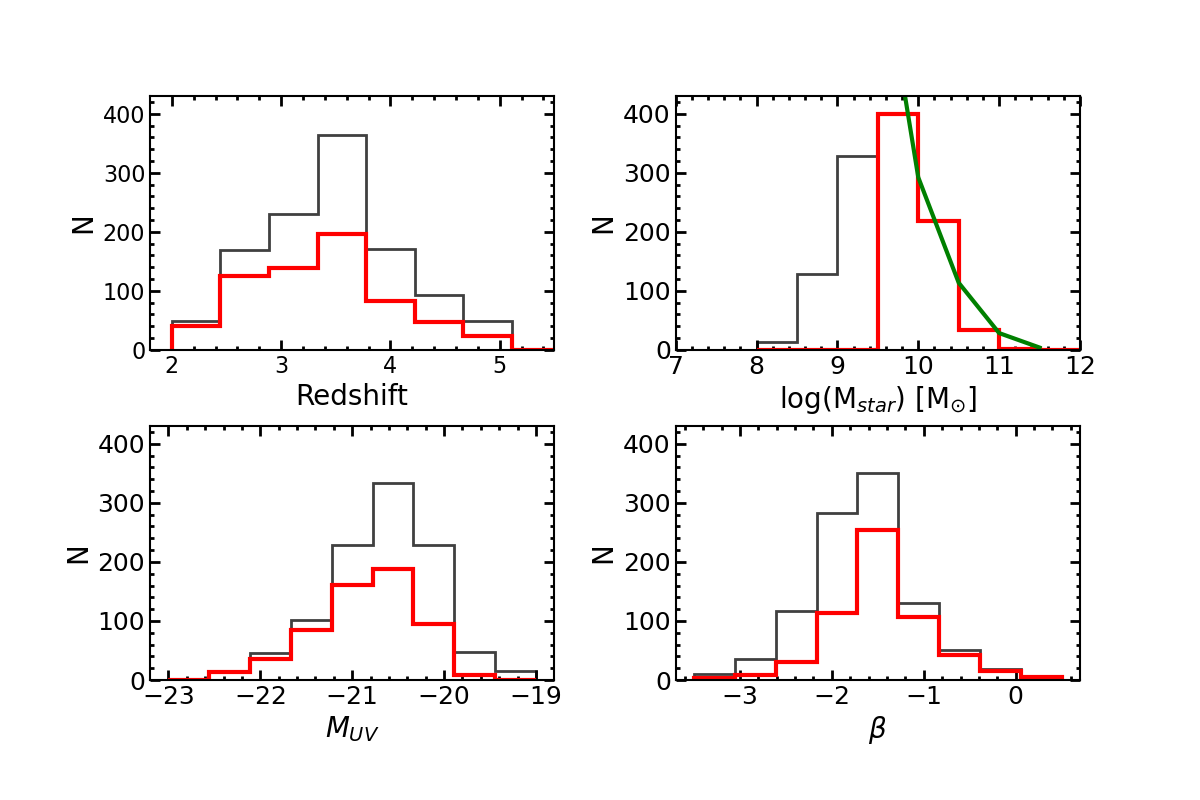}
\caption{Main properties of the VANDELS sample (black) and of the objects with log(M$_{star}$/M$_{\odot}$)$>$9.5 (red). The spectroscopic redshifts are from VANDELS DR4, stellar masses (M$_{star}$) have been estimated with \texttt{BEAGLE} as discussed in Sect.~\ref{sec:BEAGLE}, while UV magnitudes ($M_{UV}$) and slopes ($\beta$), have been measured from the observed photometry (Sect.~\ref{subsec:photometry}). The green continuous line in the top-right panel shows the M$_{star}$ distribution for all CANDELS objects at z=2-5 in UDS and GOODS-South from \citet[][]{Santini2015}, scaled by a factor of four to take into account the target sampling of the spectroscopic survey. } \label{fig_histproperties}
\end{figure} 

\begin{figure}
\centering
\includegraphics[trim={0.5cm 1.5cm 1.5cm 1.5cm},clip,width=8cm,keepaspectratio]{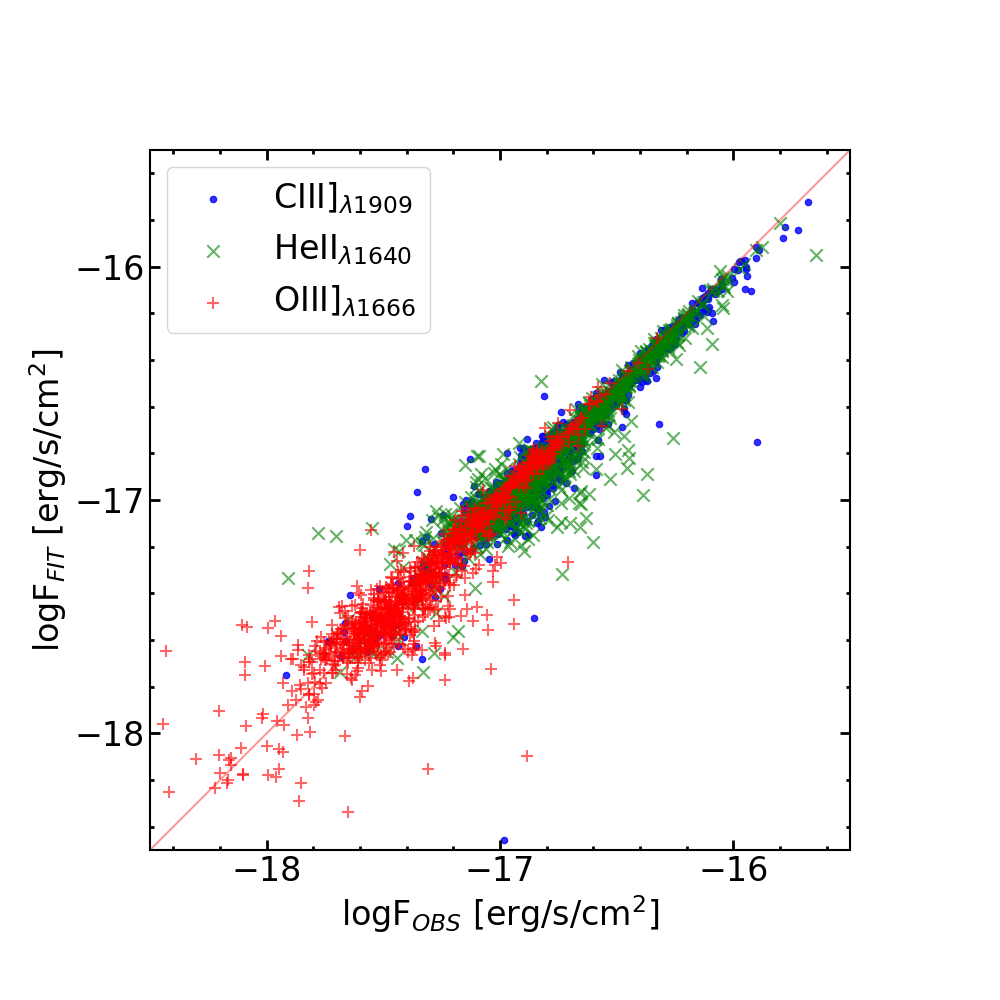}
\includegraphics[trim={0.5cm 1.5cm 1.5cm 1.5cm},clip,width=8cm,keepaspectratio]{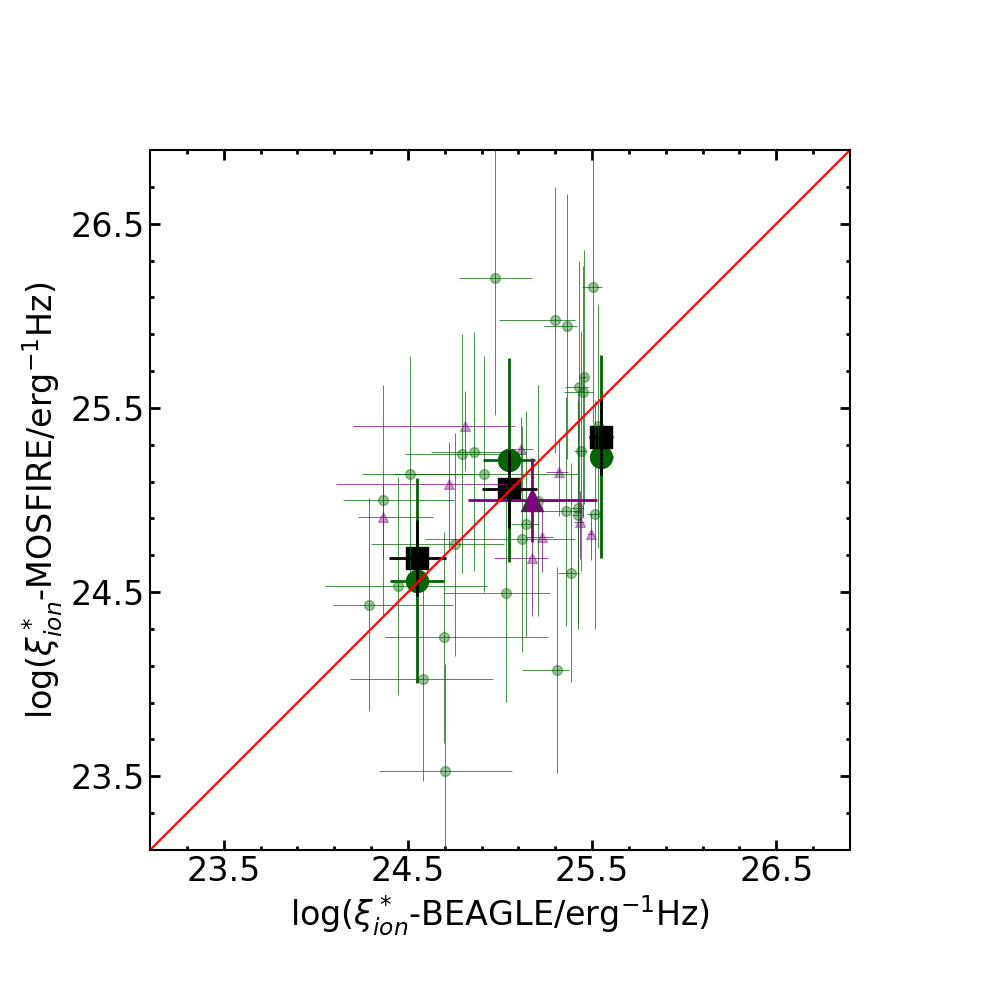}
\caption{\textbf{Top:} comparison between the \texttt{BEAGLE} best-fit flux and the observed flux of the lines used in the spectro-photometric fitting. \textbf{Bottom:} comparison between the \xion~estimated by \texttt{BEAGLE} and the measurement based on optical emission lines for the NIRVANDELS objects observed with MOSFIRE. Measurements based on the EW([OIII])-\xion~relation by \citet[][]{Chevallard2018} or on H$\alpha$ luminosity are shown as green circles and magenta triangles, respectively, with larger symbols and error bars indicating the relevant median and dispersion. The black filled squares and errorbars show the median and dispersion in bins of $\Delta$ log(\xion-BEAGLE)=0.5 for all objects with either H$\alpha$ or [OIII] measurement.} 
\label{fig_checks}
\end{figure}

\section{Spectro-photometric fitting with BEAGLE}\label{sec:BEAGLE}

We measure the physical parameters of the VANDELS galaxies, including the ionizing budget, by means of a spectro-photometric fit performed with the \texttt{BEAGLE} tool \citep{Chevallard2016} using the most recent version of the \citet[][]{Bruzual2003} stellar population synthesis models \citep[see][for details]{VidalGarcia2017}. Nebular emission is modeled self-consistently as described in \citet[][]{Gutkin2016} by processing stellar emission with the photoionization code \textsc{CLOUDY} \citep[][]{Ferland2013}. The fit is performed by fitting the integrated lines plus continuum fluxes measured as described in Sect.~\ref{sec:sample}. The redshift is fixed at the spectroscopic value, which is the one provided with the VANDELS final release, with the exception of the objects with a \ciii~detection at SNR$>$3, for which we use the relevant systemic redshift determination \citep[][]{Calabro2022a}. 
The \texttt{BEAGLE} SED-fitting runs are performed with a configuration similar to the one adopted to estimate the ionizing efficiency of very high-redshift galaxies \citep{Stark2017,Castellano2022a}. The templates are based on a \citet{Chabrier2003} initial mass function and have metallicity in the range $-2.2 \leq log(Z/Z_{\odot}) \leq 0.25$. The configuration adopts the most flexible parametric star formation history (SFH) allowed by the code, i.e, an exponentially delayed function (SFR(t) $\propto$ t$\cdot$exp(-t/$\tau$)) plus an ongoing constant burst.  This SFH model allows the analysis of objects with both rising and declining star formation histories \citep[][]{Carnall2019}, and an accurate estimates of the global properties of both main-sequence and starburst galaxies \citep[][]{Ciesla2017}.

The duration of the final constant SFR phase is a free parameter in \texttt{BEAGLE} that we chose to fix to 10 Myr considering that most of the nebular emission is generated by reprocessed light of massive stars with age of 3-10 Myr \citep[][]{Kennicutt1998,Kennicutt2012}. We adopt uniform priors on the SFH exponential timescale (7.0 $\leq$ $\tau$/log(yr) $\leq$ 10.5), stellar mass (7.0 $\leq$ log($M/M_{\odot}$) $\leq$ 12), star-formation rate (0.0 $\leq$ log($SFR/M_{\odot}/yr$) $\leq$ 3.0), and maximum stellar age (7.0 $\leq$ log($Age/yr$) $\leq$  ~age of the universe). Attenuation by dust is treated following the \citet{Charlot2000} model combined with the \citet{Chevallard2013} prescriptions for geometry and inclination effects, assuming an effective V-band optical depth in the range -3.0 $\leq$ log($\tau_{V}$) $\leq$ 0.7 with a fixed fraction $\mu=0.4$ arising from dust in the diffuse ISM. Interstellar metallicity $Z_{ISM}$ is assumed to be identical to the stellar one, the dust-to-metal mass ratio and ionization parameter are left free in the ranges $0.1 \leq \xi_d \leq 0.5$ and -4.0 $\leq$ log($U_{s}$) $\leq$ -1.0, respectively.
In the following analysis we will use the best-fit parameters obtained by \texttt{BEAGLE} from the maximum of the posterior probability distribution functions. The 68\% confidence level uncertainty on each parameter is measured from the relevant marginal probability distribution. For the derived parameter \xion, which is not expressly sampled over in the fitting, we provide the mean and 68\% confidence level interval of the relevant weighted distribution derived from the \texttt{MULTINEST} samples (Equation 9 in \citealt{Feroz2008}, see also \citealt{Chevallard2016}).

We show in Fig.~\ref{fig_histproperties} the main properties of the sample. The VANDELS objects are mostly bright ($M_{UV}<$-20), massive galaxies whose redshift distribution peaks at z$\sim$3-4 with tails covering the entire z$\sim$2-5 range. The stellar mass distribution covers the range log(M$_{star}$/M$_{\odot}$)$=$8-11, but it is clearly incomplete at low masses. We compared the mass distribution of the VANDELS sample to the one of all objects with photometric redshift $2<z<5$ in the official CANDELS catalogs \citep[][]{Santini2015}: the two are perfectly consistent at log(M$_{star}$/M$_{\odot}$)$>$9.5 once the CANDELS counts are scaled by a factor of four to take into account the target sampling of the spectroscopic survey. Instead, the VANDELS target selection criteria clearly yields to incompleteness at lower masses. For this reason in the following we will discuss separately the entire sample and the "mass-complete" subsample at log(M$_{star}$/M$_{\odot}$)$>$9.5.

\subsection{Accuracy of the spectro-photometric analysis}
Our approach exploits a combined spectro-photometric fit to estimate \xion~and other physical parameters of the galaxies at the same time. As a test of the reliability of the fit including spectroscopic lines we compared the measured line fluxes to the relevant best-fit fluxes from \texttt{BEAGLE}: as shown in Fig.~\ref{fig_checks} \texttt{BEAGLE} can accurately reproduce the three observed lines across the entire flux range probed. 

We then assessed the dependence of our results on the adopted \texttt{BEAGLE} configuration through the analysis of a representative subsample of 200 objects at z=3-4. In order to test the effect of different attenuation laws we performed the fit adopting the \citet[][]{Calzetti2000} one or the SMC extinction law by \citet[][]{Pei1992}. We find that the \xion~values from our reference run which assumes a \citet[][]{Charlot2000} attenuation are consistent ($\Delta$ log(\xion)$<$0.01)  with those found with a \citet[][]{Calzetti2000} law, but on average slightly lower ($\Delta$ log(\xion)=0.065) than those based on the SMC extinction.

We then tested the dependence of the \xion estimates on the assumed SFH. We find a that the spectrophotometric fit returns slightly higher \xion values \citep[$\sim$0.1 dex, see also][]{Mascia2023}  when assuming a 5 Myr instead of a 10 Myr timescale as duration of the ongoing star-formation episode.
We then performed a fit using only the photometric information and setting the duration of the ongoing star-formation episode to 100 Myr matching the SFR timescale probed by UV integrated light. We found an almost fixed \logxion$\sim$25.2 for all sources, while the SFR and $M_{star}$ values are consistent with the values in the reference spectrophotometric fit, albeit with a fraction of objects having SFR lower by a factor 1.5, and correspondingly higher stellar mass. Instead, a fit using photometry only and a 10Myr duration of the ongoing star-formation episode provided \xion~values statistically consistent with those obtained for the same SFH and including spectroscopic information, with the exception of a fraction of objects with EW(\oiii)$>$3 for which the ionizing efficiency is found to be lower. This result indicates that the inclusion of spectroscopic measurements enables more accurate constraints although the estimated efficiencies are largely affected by photometric information in objects with a low EW of the UV lines. 

In order to perform a direct test of the robustness of our \xion~measurement we exploited observation from the NIRVANDELS survey that targeted the VANDELS fields with MOSFIRE to measure optical rest-frame emission lines. A detailed description of the NIRVANDELS observations can be found in \citet[][]{Cullen2021}. We found that 30 objects at z$\sim$3.15-3.78 in our sample have clear [OIII]$\lambda 4959,5007$ detection in their MOSFIRE K-band spectra enabling an estimate of \xion~from EW(OIII) using Eq.(3.1) in \cite{Chevallard2018}. In addition, 9 objects at z$\sim$2.23-2.49 have detection of both $H\alpha$ and $H\beta$, enabling a measurement of \xion~using standard conversions from the dust-corrected H$\alpha$ luminosity  \citep[e.g.,][]{Shivaei2018}. We measure the total flux of the Balmer lines and of the [OIII] $\lambda \lambda$4959,5007 doublet with a Gaussian fit of each line component.
For the z$>$3 subsample we use the measured broad band photometry at the corresponding observed wavelength to determine the EW([OIII]) since the continuum is undetected in the spectra. For the objects with detection of the Balmer lines we first correct the measured $H\alpha$ luminosity for dust extinction on the basis of the Balmer decrement assuming a \citet[][]{Calzetti2000} attenuation law and an intrinsic ratio $(H\alpha/H\beta)=2.86$ \citep[see, e.g.,][]{Dominguez2013}. We then apply the relation in \citet[][]{Leitherer1995} to convert $L(H\alpha)$ into an intrinsic Lyman-continuum photons production rate $N(H^0)$ and we estimate the ionizing efficiency as $\xi_{ion}^{*}=N(H^0)/L_{UV}$, where $L_{UV}$ is the dust-corrected UV luminosity. We have assumed a null escape fraction of ionizing photons, considering that this quantity is highly uncertain but most likely lower than 3-5\% for bright, massive objects in this redshift range \citep[e.g.][]{Grazian2016,Begley2022}. We also
 neglected corrections for Balmer absorption which has been found to be very small \citep[$\sim$3\% on average on NIRVANDELS sources,][]{Cullen2021}. The comparison between the \xion~estimated by \texttt{BEAGLE} and the one measured from optical emission lines is shown in Fig.~\ref{fig_checks}. We find a large scatter on individual measurements, which is however consistent with the relevant uncertainties. Most importantly, we find a fair agreement ($\Delta$ log(\xion)$<$0.1) on average among the different \xion~estimates, with no evident systematic trends. We also find agreement on average when analysing separately the two subsamples of $z>3$ [OIII] emitters and of $z\sim2$ $H\alpha$ emitters. The former subsample is large enough to assess consistency as function of \xion. In turn, the individual measurements based on Balmer lines do not show a clear trend, but the low number of objects in this subsample only allows to assess consistency of the global median value. While it is advisable to perform more in-depth tests of this kind with future spectroscopic samples, we conclude that the \xion~estimates obtained by our spectro-photometric fitting procedure are reliable for our purpose of exploring correlations between the ionizing efficiency and the physical properties of the galaxies in the VANDELS sample.

\begin{figure}
\centering
\includegraphics[trim={1.8cm 0.7cm 1.5cm 0.05cm},clip,width=\linewidth,keepaspectratio]{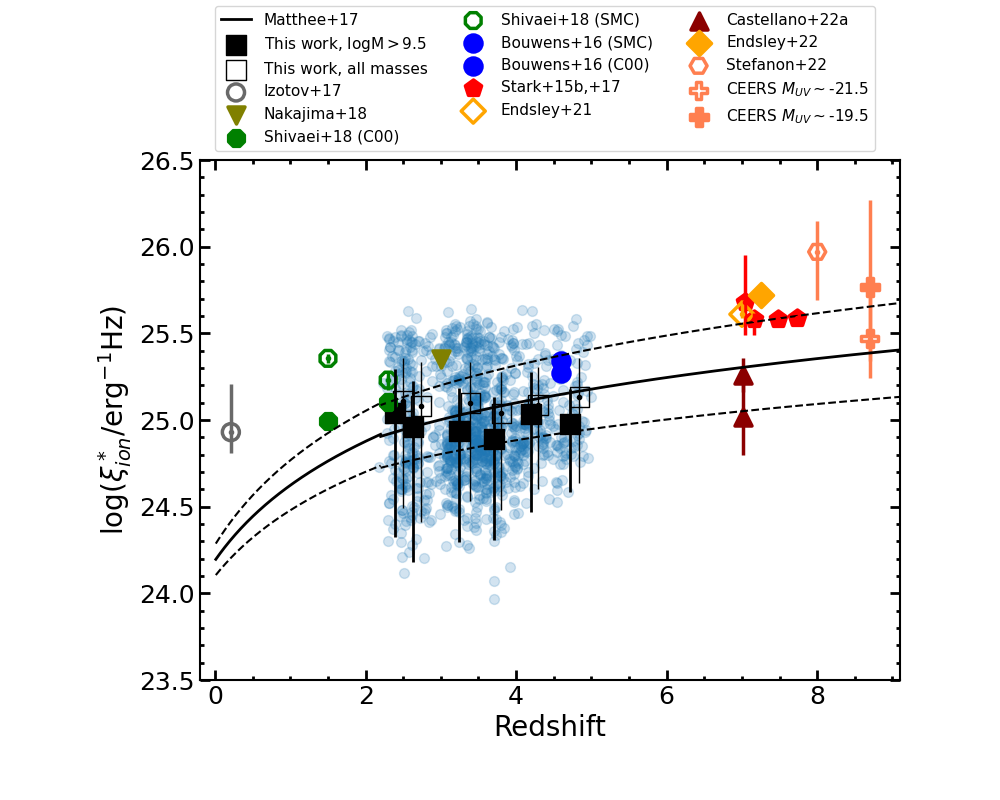}
\caption{\xion~as a function of redshift: black empty and filled squares and error bars show the average and standard deviation for all VANDELS objects and for those with log(M$_{star}$/M$_{\odot}$)$>$9.5, respectively. Individual VANDELS sources are shown as blue circles. The VANDELS measurements are compared to the observed \xion~versus redshift relation by \citet[][]{Matthee2017} (black continuous and dashed lines show average and dispersion, respectively), and to measurements from the literature for SDSS compact star-forming galaxies \citep[][]{Izotov2017}, LBGs at z$\sim$2-3 \citep[][]{Shivaei2018} and z$\sim$4-5 \citep[][]{Bouwens2016a}, star-forming galaxies at z$\sim$2-4 \citep[][]{Nakajima2018a}, individual objects at z$\sim$7-8 \citep[][]{Stark2015b,Stark2017,Castellano2022a}, bright \citep[][]{Endsley2021a} and faint \citep[][]{Endsley2022c} LBGs at z$\sim$7, LBGs at $z\sim 8$ \citep[][]{Stefanon2022}, bright and faint LBGs at $z\sim 8-9$ observed with JWST-NIRSpec by the CEERS ERS survey \citep[][]{Fujimoto2023}. The measurements by \citet[][]{Shivaei2018} and \citet[][]{Bouwens2016a} are shown for both the SMC extinction law and the \citet[][]{Calzetti2000} attenuation law. See legend for symbols and colors. } \label{fig_vs_redshift}
\end{figure} 

\begin{figure}
\centering
\includegraphics[trim={0.5cm 0.3cm 0.5cm 0.5cm},clip,width=\linewidth,keepaspectratio]{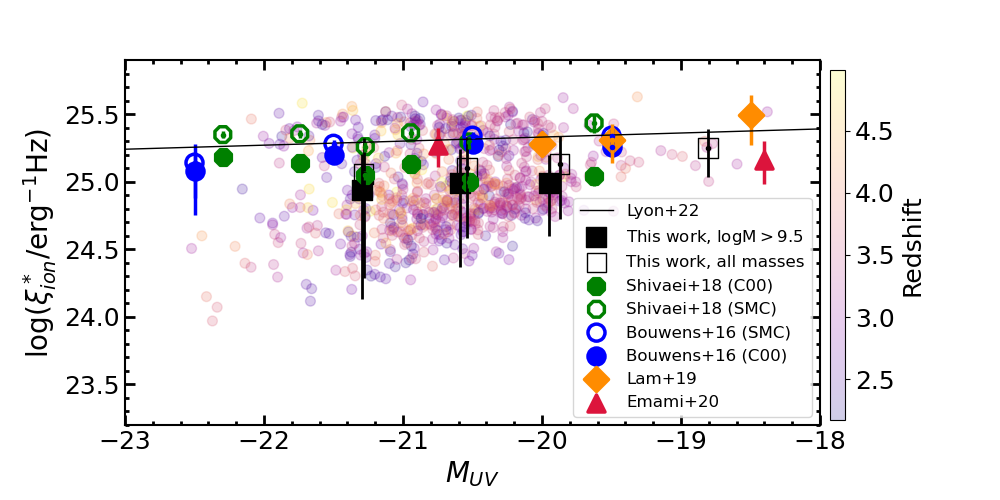}
\includegraphics[trim={0.5cm 0.3cm 0.5cm 0.5cm},clip,width=\linewidth,keepaspectratio]{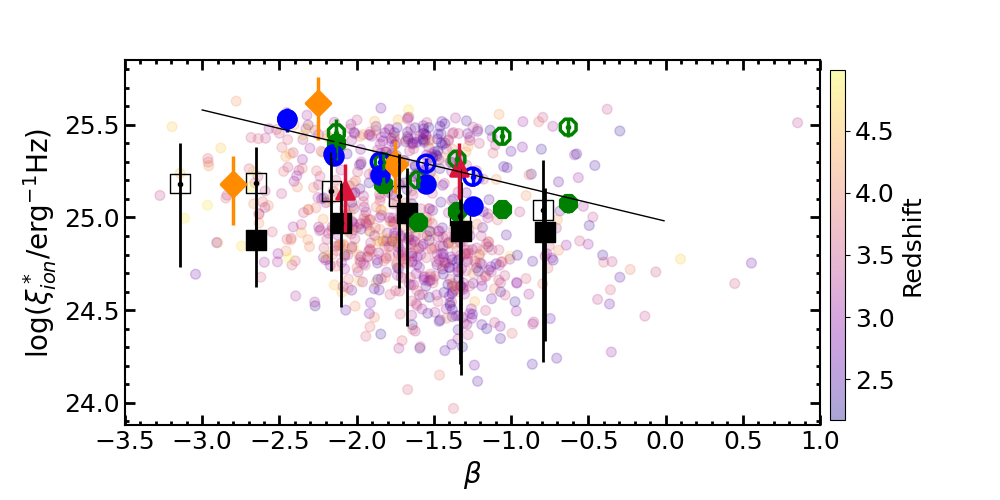}
\caption{\xion~as a function of $M_{UV}$ (top) and  UV slope (bottom). The VANDELS objects are color-coded according to the relevant redshift. Black empty and filled squares and error bars show the average and standard deviation for the entire sample and for galaxies with log(M$_{star}$/M$_{\odot}$)$>$9.5, respectively. The VANDELS sample is compared to measurements from the literature by \citet[][]{Shivaei2018} (LBGs at $z\sim$2-3), \citet[][]{Emami2020} (lensed dwarf galaxies at 1.4$<z<$2.7), \citet[][]{Bouwens2016a} and \citet[][]{Lam2019} (LBGs at $z\sim$4-5). See legend for symbols and colors. The black continuos lines show the relations measured by \citet[][]{Prieto-Lyon+22} on sub-$L^*$ galaxies at $z\sim3-7$.} \label{fig_vs_MUVbetaLya}
\end{figure} 

\section{Results}\label{sec:results}
The combined spectro-photometric constraints on the VANDELS sample allow us to explore correlations between \xion~and all properties of interest for high-redshift populations to search for the most reliable indicators of a high-ionizing efficiency. The average and standard deviation of \xion~as a function of the quantities discussed below are reported in Tables~\ref{tab:obsproperties} and ~\ref{tab:physproperties} for the reference sample with log(M$_{star}$/M$_{\odot}$)$>$9.5. Throughout the paper we will discuss the potential relation between \xion~and other galaxy properties on the basis of the relevant \citet[][]{Spearman04} rank coefficients measured with the \texttt{spearmanr} algorithm from the \texttt{scipy} library. The Spearman rank test assesses whether a monotonic relation exist between two variables, without any assumption on the form of the relation. The rank coefficient is defined in the range $-1<r_s<1$, where negative (positive) values indicate anti-correlation (correlation) between the two variables. The relevant p-value $p(r_s)$ is the probability of the null hypothesis of absence of any correlation. We will consider a correlation to be present whenever $p(r_s)<$0.01. The Spearman rank coefficients will be summarised in Fig.~\ref{fig_spearman} both for the entire sample and the high-mass one. In the forthcoming sections we will show the correlations estimated from the VANDELS sample together with available results from the literature. A detailed comparison with previous works will be discussed in Sect.~\ref{sec:comparison}, but we anticipate here that differences in the stellar mass distributions between VANDELS and other samples likely explain discrepancies found with respect to other observed and physical properties.

\subsection{Dependence of \xion~on observed properties}\label{subsec:obsproperties}
We first explore the relation between \xion~and observed photometric and spectroscopic properties of the VANDELS sample. The ionizing efficiency as a function of redshift is shown in Fig.~\ref{fig_vs_redshift} together with the relevant average values in $\Delta z$=0.5 bins. The objects span a large \xion~range, with no significant evolution with redshift ($r_s\sim$0.05). 

We show in Fig.~\ref{fig_vs_MUVbetaLya} the relations between \xion and $M_{UV}$, and between \xion and UV slope $\beta$, color-coding each object according to the relevant redshift.
We find mild, albeit significant correlations in both cases.
The ionizing efficiency increases at fainter $M_{UV}$ ($r_s$=0.23) from 24.8 at $M_{UV}\sim$-21.4, to $\sim$25.2 at at $M_{UV}\sim$-18.8. Similarly, \xion~increases at decreasing UV slope $\beta$ ($r_s$=-0.23). These trends are weaker, although still significant according to the Spearman correlation test, when considering only objects with log(M$_{star}$/M$_{\odot}$)$>$9.5. In fact, the relation with UV slope is nearly flat with the possible exception of objects at extremely blue $\beta<$-3. No clear redshift-dependent effect is evident in the above mentioned diagrams, with objects at z$\sim$2-5 populating all the observed range.

\begin{figure}
\centering
\includegraphics[trim={3cm 1.5cm 0.8cm 1.8cm},clip,width=\linewidth,keepaspectratio]{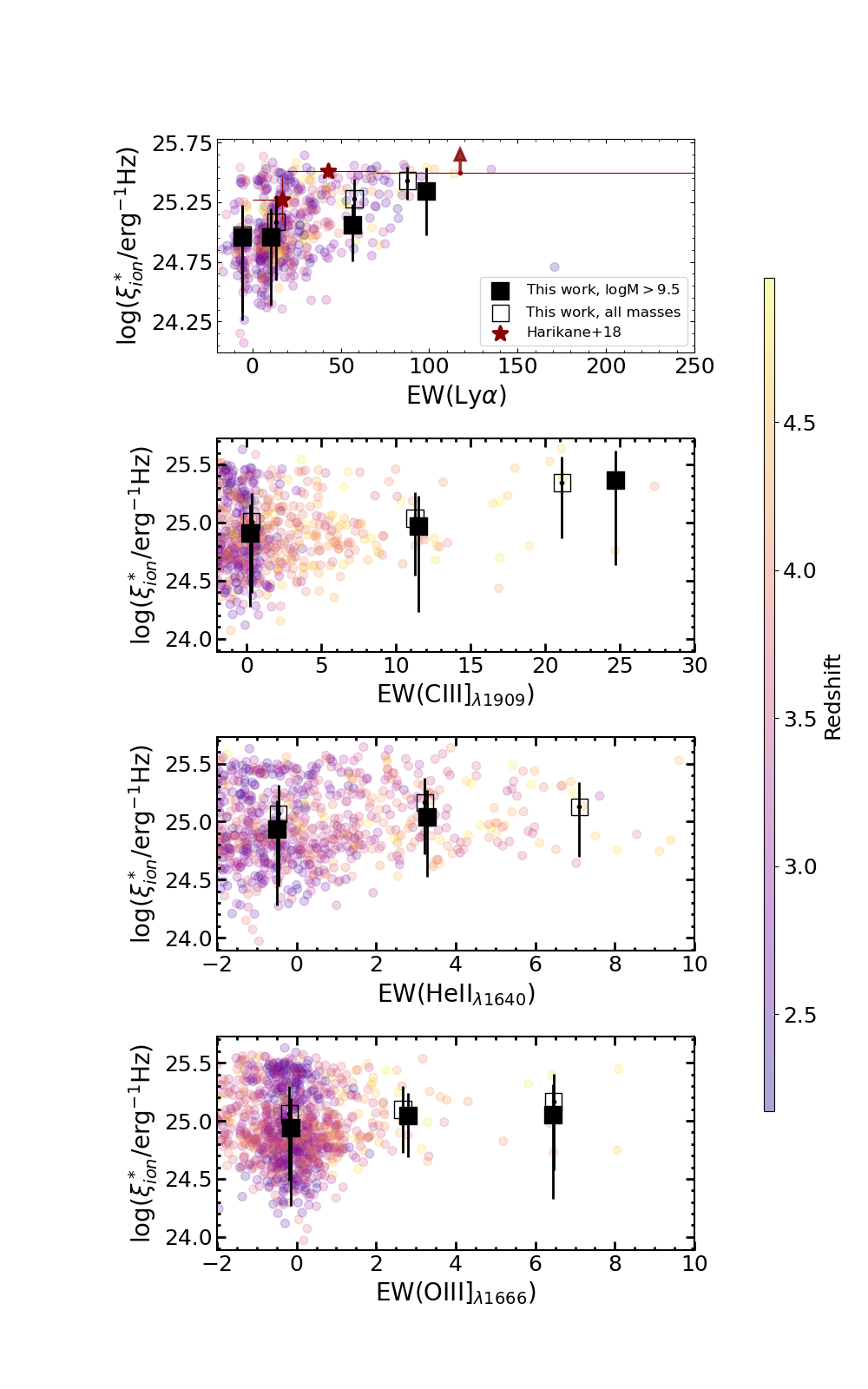}
\caption{Same as Fig.~\ref{fig_vs_MUVbetaLya} but for, from top to bottom, the EWs of Ly$\alpha$, CIII], HeII and OIII]. Negative values on single objects are due to absorption in the case of Ly$\alpha$, and to measurement noise on galaxies with EW$\sim$0 for the other lines. The  \xion-EW(Ly$\alpha$) in the top panel is compared to measurements on LAEs at z$\sim$5 by \citet[][]{Harikane2018}} \label{fig_vs_UVlines}
\end{figure}

The relation between ionizing efficiency and the EW of the UV emission lines is shown in Fig.~\ref{fig_vs_UVlines}. There is a robust, significant trend of increasing \xion at increasing EW(Ly$\alpha$) ($r_s$=0.42) with an average \logxion$>$25 at EW$>$50\AA. Instead, the correlations between \xion and other emission lines are much milder. According to the Spearman test the  correlations between \xion~and the EW of both \ciii~($r_s$=0.22) and \oiii~($r_s$=0.27) are significant when considering objects in the full sample having a line detection, but only the latter remains significant (with $r_s$=0.36) for the objects with log(M$_{star}$/M$_{\odot}$)$>$9.5. The positive correlation with EW(\ciii) is likely driven by few objects having \ciii~with EW$>$20\AA~and average \logxion$\sim$25.2. In turn, no significant correlation is found between \xion~and the EW of \heii.

\subsection{Dependence of \xion~on physical parameters}\label{subsec:fitproperties}
\begin{figure}
\centering
\includegraphics[trim={0.1cm 0.1cm 0.1cm 0.1cm},clip,width=9.5cm,keepaspectratio]{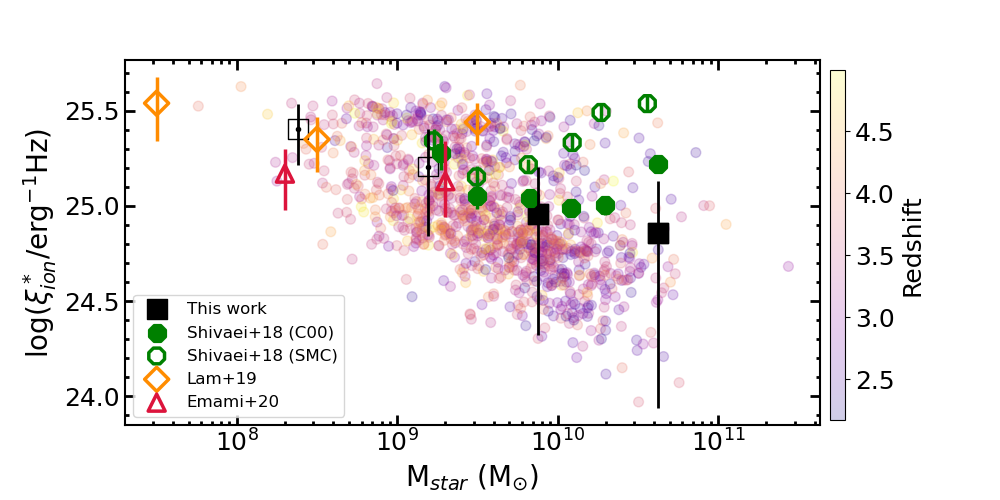}
\includegraphics[trim={0.1cm 0.1cm 0.1cm 0.1cm},clip,width=9.5cm,keepaspectratio]{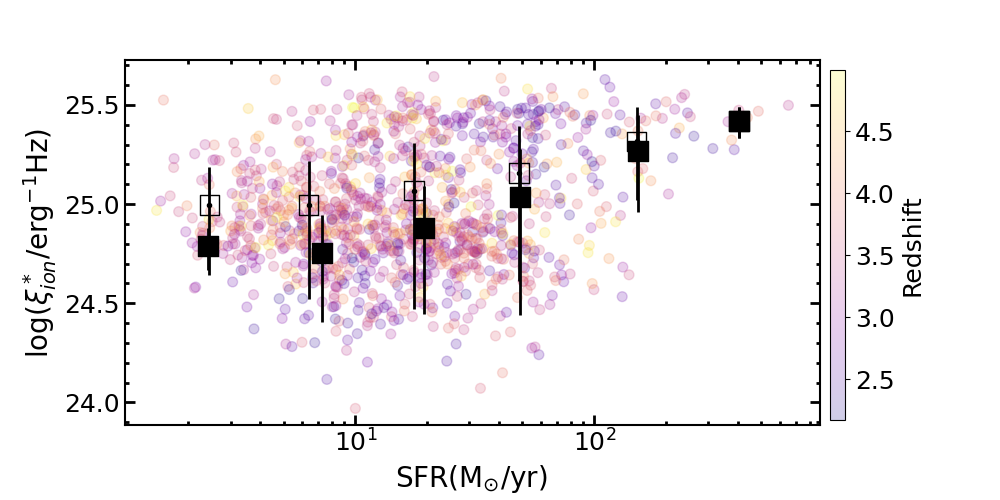}
\includegraphics[trim={0.1cm 0.1cm 0.1cm 0.1cm},clip,width=9.5cm,keepaspectratio]{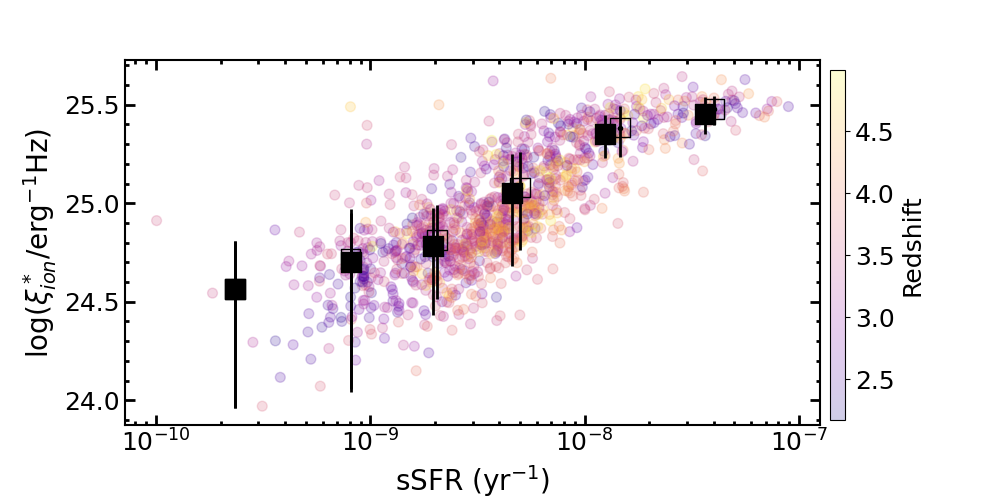}
\includegraphics[trim={0.1cm 0.1cm 0.1cm 0.1cm},clip,width=9.5cm,keepaspectratio]{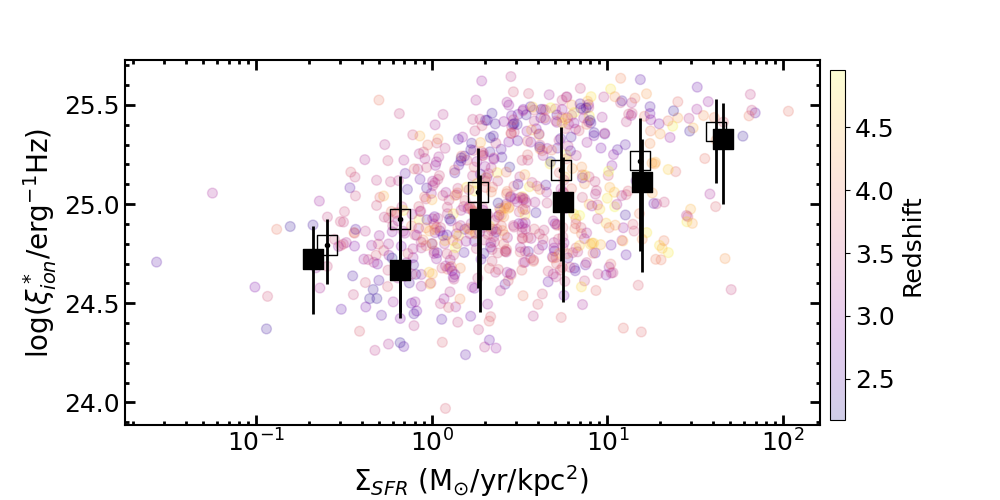}
\caption{Same as Fig.~\ref{fig_vs_MUVbetaLya} but for, from top to bottom, M$_{star}$, SFR, sSFR and $\Sigma_{SFR}$.} \label{fig_MstarSFR}
\end{figure}

In Fig.~\ref{fig_MstarSFR} we show the relation between \xion~and physical properties for the VANDELS sample. We find a significant trend of increasing average ionizing efficiency at decreasing mass ($r_s$=-0.54), with \logxion~going from $\sim$24.8 at log(M$_{star}$/M$_{\odot}$)$>$11 to $\gtrsim$25 at log(M$_{star}$/M$_{\odot}$)$<$10. A large scatter is found on the \xion-SFR plane, but there is a clear prevalence of objects with high \xion~at SFR$>$100$M_{\odot}/yr$, and the Spearman test indicate a significant correlation for both the full sample ($r_s$=0.16) and the high-mass sample ($r_s$=0.38). 

\begin{figure}
\centering
\includegraphics[trim={0.8cm 0.9cm 0.2cm 0.5cm},clip,width=\linewidth,keepaspectratio]{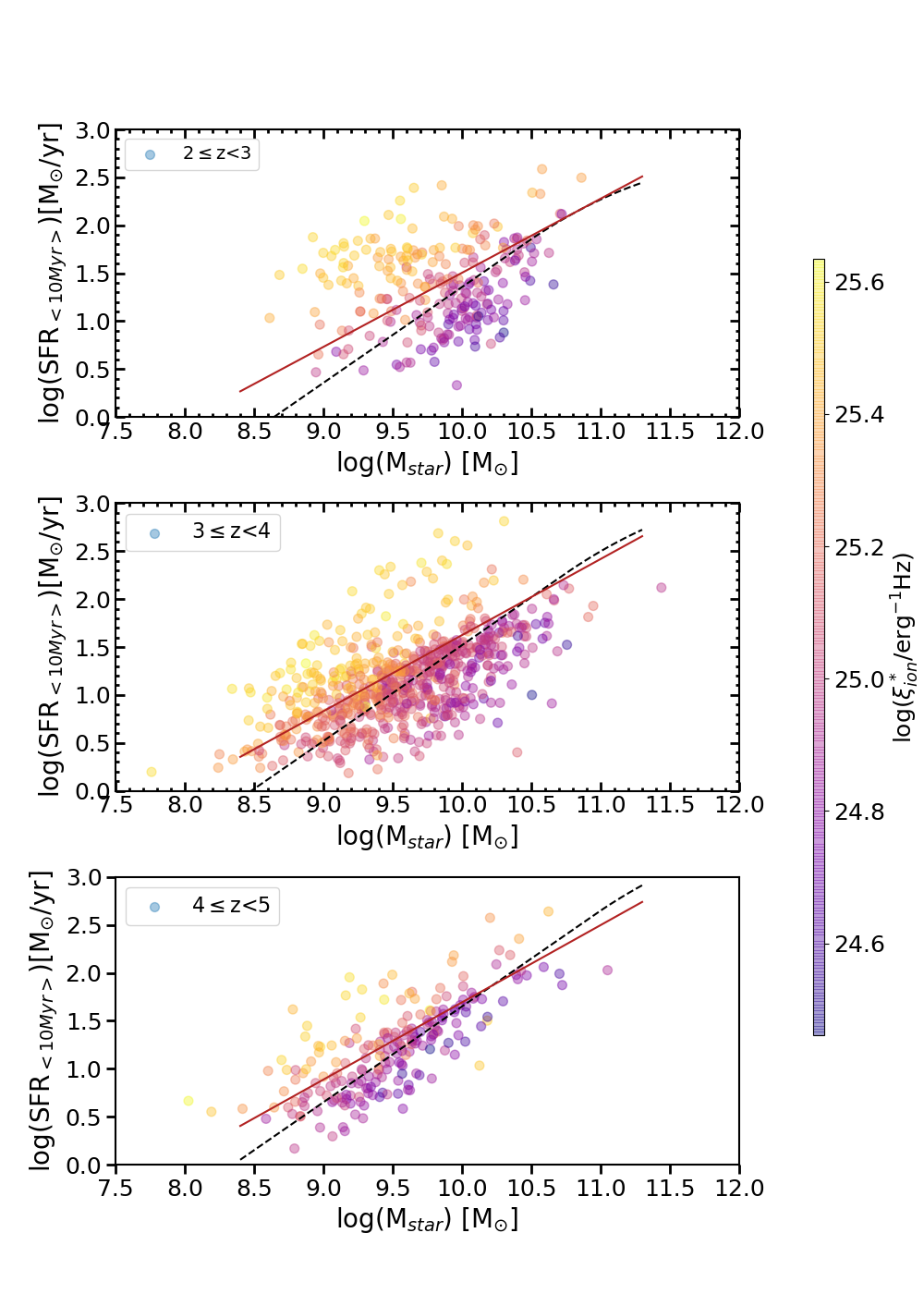}
\caption{The VANDELS sample on the SFR versus M$_{star}$ plane. Each object is color-coded according to the relevant \xion~value. The best-fit estimates of the main sequence of star-formation by \citet[][]{Speagle2014} and by \citet[][]{Schreiber2015} are shown as a red continuos line and a black dashed curve, respectively.} \label{fig_MS}
\end{figure}

The most evident correlation is found between \xion~and sSFR, with a monotonic increase from \logxion  $\sim$24.5 at log(sSFR)$\sim$-9.5$yr^{-1}$ to $\sim$25.5 at log(sSFR)$\sim$-7.5$yr^{-1}$. Most importantly, at variance with the other relations discussed here, the sSFR-\xion~relation has a  low scatter, and no evidence for multi-modal distributions.  The correlation is significant according to the Spearman test and shows little dependence on mass, with the full sample and the log(M$_{star}$/M$_{\odot}$)$>$9.5 one having $r_s$=0.79 and $r_s$=0.68 respectively, and similar average values as a function of sSFR. The correlation is particularly evident when analysing the VANDELS sample in the M$_{star}$-SFR plane (Fig.~\ref{fig_MS}): objects above the main-sequence of star-formation have higher-than-average \xion, and, viceversa low \xion~is found in galaxies that have lower SFR than the typical value at their stellar mass.

The ionizing efficiency also shows a significant increase in compact star-forming objects ($r_s$=0.35-0.38 for the full and the high-mass samples). The galaxies with $\Sigma_{SFR}>$10 M$_{\odot}/yr/kpc^2$ having \logxion$>$25 with very few exceptions, and such high efficiencies being prevalent at $\Sigma_{SFR}>$1 M$_{\odot}/yr/kpc^2$.

Building on the results discussed above, we searched for an equation to provide an estimate of \xion on the basis of the galaxy physical properties. To this aim, we carried out a fully data-driven power regression analysis to evaluate equations combining the measured parameters following the approach described in \citet{Mascia2023b}. A regularized minimization of the sum of the root-mean-squared error (RMSE) and of the mean absolute error (MAE), computed between the values provided by each equation and the dataset, yields the following regression as best description of our dataset: 

\begin{center}
\xion = $c_0$ + $c_1$ ${\log(\text{SFR})}^{\gamma_{1}}$ + $c_2$ ${\log(M_{star}/M_{\odot})}^{\gamma_{2}}$,
\end{center}

The coefficients and corresponding uncertainties have been estimated by repeating the minimization process 1000 times with a bootstrap approach where in every iteration 10\% of the sample was randomly removed:

\begin{center}
$c_0 = 0.002 \pm 0.010$,\\
$c_1 = 0.329 \pm 0.015$,\\
$c_2 = 38.63 \pm 0.21$,\\
$\gamma_{1} = 1.28 \pm 0.04$,\\
$\gamma_{2} = -0.200 \pm 0.003$.\\
\end{center}

The power law model presented above yields to a RMSE=0.2 and a MAE=0.15. The fact that this particular model, which is a function of the SFR and the stellar mass, was selected as the best alternative to minimize the error is consistent, from a physical perspective, with the fact that a clear correlation is observed between the sSFR and \xion.

\begin{figure}[t]
\centering
\includegraphics[trim={0.1cm 0.1cm 0.1cm 0.1cm},clip,width=\linewidth,keepaspectratio]{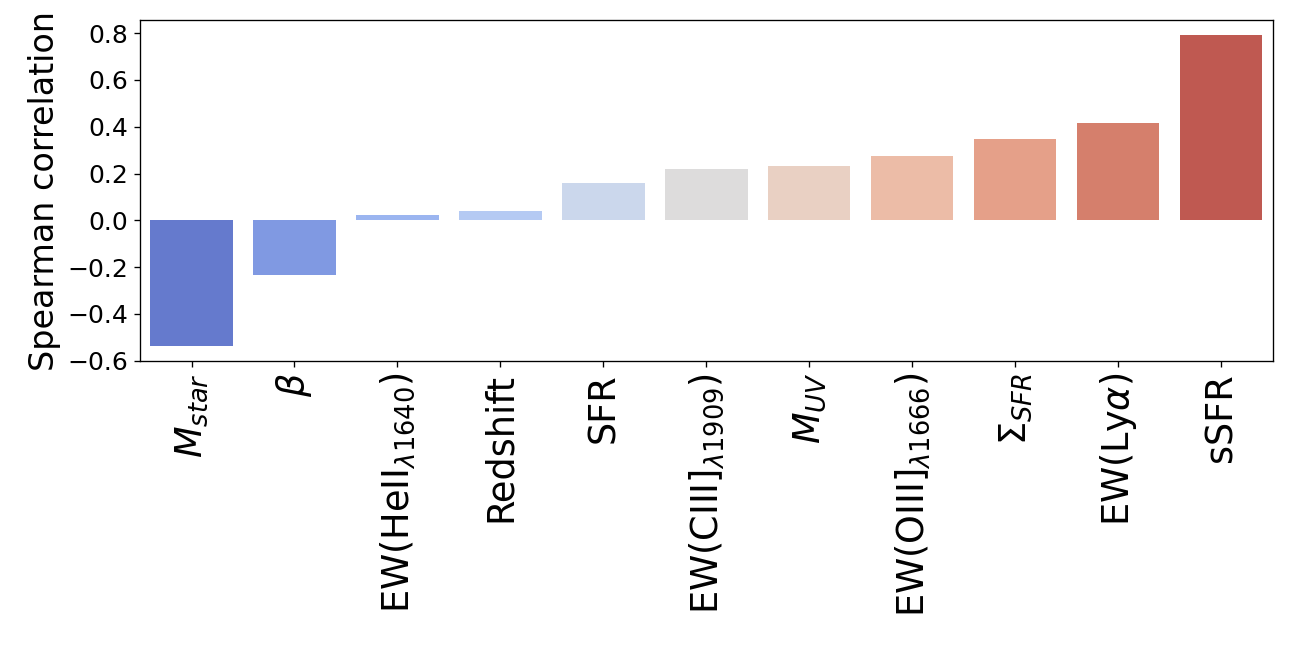}
\includegraphics[trim={0.1cm 0.1cm 0.1cm 0.1cm},clip,width=\linewidth,keepaspectratio]{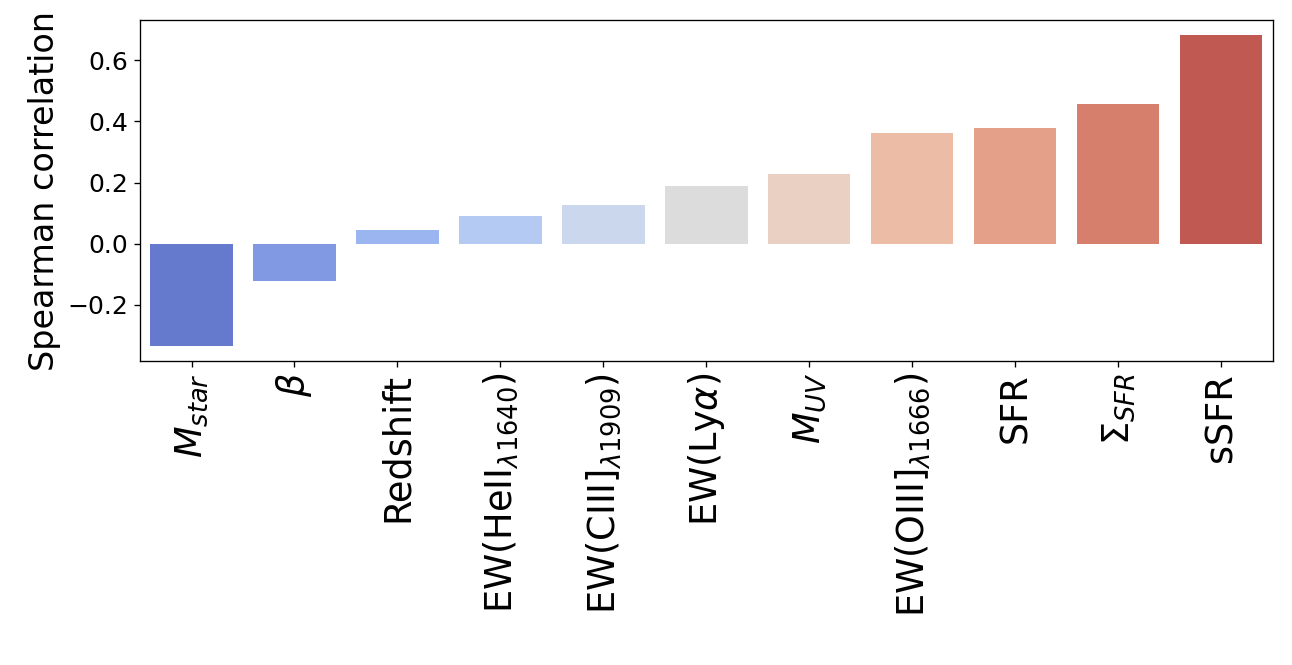}
\caption{Spearman rank coefficients for correlation between \xion~and the various properties analysed in this paper, for the entire sample (top) and for the log(M$_{star}/M_{\odot})>9.5$ subsample (bottom).} \label{fig_spearman}
\end{figure} 

\begin{figure}
\centering
\includegraphics[trim={1cm 1cm 1cm 1cm},clip,width=\linewidth,keepaspectratio]{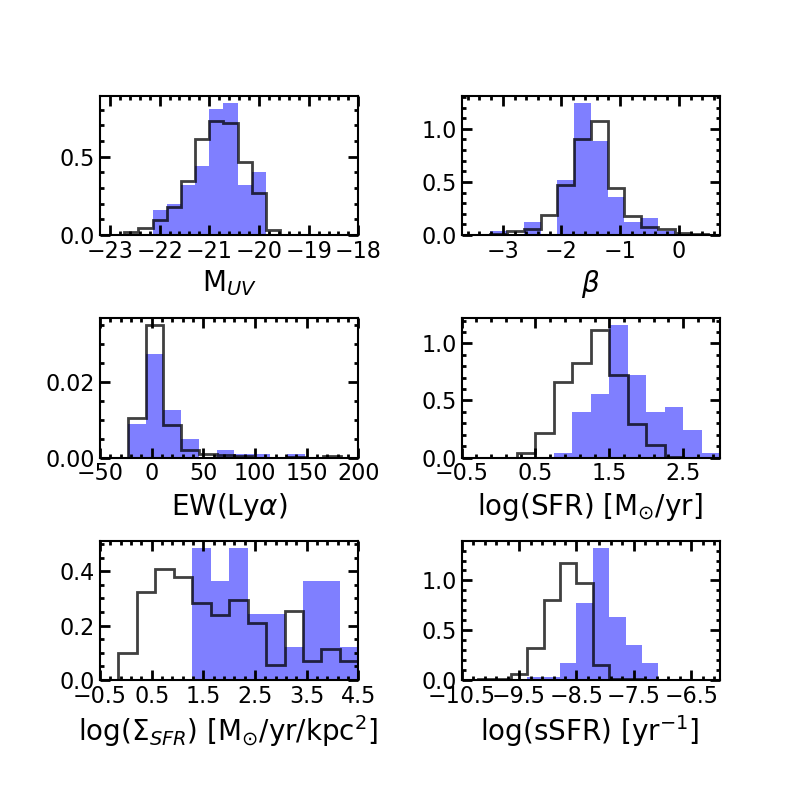}
\caption{Probability density distributions of, from top left to bottom right, $M_{UV}$,  UV slope, EW(Ly$\alpha$), SFR, $\Sigma_{SFR}$ and sSFR for objects with \xion~$>$25.2 (blue filled histogram) and objects with \xion~$<$25.2 (black empty histogram). Only galaxies with log(M$_{star}/M_{\odot})>9.5$ are considered in both cases.} \label{fig_ionizers}
\end{figure}

\subsection{Comparison with previous results}\label{sec:comparison}
The relation between \xion~and redshift as well as its correlation with observed properties have been explored in several recent works. In Fig.~\ref{fig_vs_redshift} we compare our estimates to previous measurements of \xion from representative samples of star-forming galaxies at different redshifts. The average ionizing efficiency as a function of redshift computed on the entire VANDELS sample (black empty squares in Fig.~\ref{fig_vs_redshift}) is in agreement with the \xion-redshift relation by \citet[][]{Matthee2017}, while our log(M$_{star}$/M$_{\odot}$)$>$9.5 sample (black filled squares) lies below their average. The average \xion value measured by \citet[][]{Nakajima2018a} on a composite of star-forming galaxies at $z\sim$2-4 is higher than both the VANDELS and the \citet[][]{Matthee2017} averages, but the difference can be partly due to their use of \texttt{BPASS} models including binary stellar populations with a 300M$_{\odot}$ upper-mass cut-off of the IMF \citep[][]{Stanway2016}.

Our objects also show lower average \xion~than the samples by \citet[][]{Shivaei2018} and \citet[][]{Bouwens2016a} at z$\sim$2 and z$\sim$4.5, respectively, in particular when comparing to estimates based on a Small Magellanic Cloud (SMC) extinction law. 
Similarly, the \xion-$M_{UV}$, and \xion-$\beta$ relations in our sample have lower average values than the ones from the literature \citep[][]{Bouwens2016a,Lam2019,Emami2020,Prieto-Lyon+22}, although, as discussed in Sect.~\ref{subsec:obsproperties}, both relations appear bimodal, and the locus of VANDELS objects with higher ionizing efficiency includes all observed values from the literature. Our average \xion~values at $M_{UV}\lesssim$-20, and as a function of UV slope are in better agreement with the measurements by \citet[][]{Shivaei2018} based on the \citet[][]{Calzetti2000} attenuation law.  

Most importantly, a good agreement is found with the \xion-M$_{star}$ relations by both \citet[][]{Shivaei2018} (based on Calzetti attenuation), \citet[][]{Lam2019} and \citet[][]{Emami2020} at log(M$_{star}$/M$_{\odot}$)$<$10. At higher stellar masses the VANDELS sample has a lower average ionizing efficiency compared to the $1.4 \leq z \leq 3.8$ sample by \citet[][]{Shivaei2018} whose \xion-M$_{star}$ relation bends and increases again at very high masses. Considering the apparent bimodal structure of the \xion-M$_{star}$ distribution and the low number of VANDELS objects in the most massive bin shown in Fig.~\ref{fig_MstarSFR}, the difference at high masses can be explained by sample variance or by the different redshift range analysed.

The comparison against previous works on the \xion-M$_{star}$ plane is extremely important to explain the above mentioned discrepancies. In fact, the tendency towards lower \xion~values compared to \citet[][]{Bouwens2016a} and other works is most likely explained by the different mass distributions in the two samples, with an average log(M$_{star}$/M$_{\odot}$)$\sim$9.2 in their sample compared to $\sim$9.9 in VANDELS. This is also evident for the comparison with the results by \citet[][]{Lam2019}, whose sample extends to log(M$_{star}$/M$_{\odot}$)$<$8. The sample by \citet[][]{Prieto-Lyon+22} includes $Ly\alpha$-detected objects with an average $M_{UV}\sim$-18 that are most likely much less massive than the VANDELS galaxies. Interestingly, the average values for the full VANDELS sample are in better agreement with the ones from \citet[][]{Lam2019} and \citet[][]{Emami2020} also on the \xion-$\beta$ plane. The lower average \xion~values in our sample are likely connected to high-mass objects being older and more metal-rich. 

The \xion-EW(Ly$\alpha$) relation in our data is also consistent with previous findings for z$\sim$5 LAEs by \citet[][]{Harikane2018} but with average values that are $\sim$0.2dex lower at fixed EW. This discrepancy can also be likely explained by our sample of bright LBGs having a higher average stellar mass, thus being most likely older and more enriched, compared to the log(M$_{star}$/M$_{\odot}$)$\sim$8-9 of the z$\sim$5 narrow-band selected LAEs, although we cannot exclude that redshift evolution in the properties of Ly$\alpha$ emitters may play a role.

A correlation between \xion~and sSFR has been previously found by \citep[][]{Izotov2021} for CSFGs at z$<$1. They constrained a clear trend between \xion~and the quantity SFR$^{-0.9}\times M_{star}$ which is monotonically correlated with the gas-phase metallicity according to the fundamental mass-metallicity relation \citep[][]{Mannucci2010}. Once recasting our \xion-sSFR relation in this plane (not shown here) we find a consistent trend but with a lower scatter and a slight lower normalization ($\sim$0.1 dex) which is again possibly due to the lack of low-mass galaxies in the VANDELS sample.

\section{Summary and discussion}\label{sec:summary}
We have used the spectro-photometric fitting code \texttt{BEAGLE} to measure the ionizing efficiency and other physical parameters for a sample of 1174 galaxies  with secure spectroscopic redshift at z$\sim$2-5 in the VANDELS survey. The sample comprises mostly bright ($M_{UV}<$-20), massive galaxies, with a high completeness at log(M$_{star}$/M$_{\odot}$)$>$9.5. The measurement of physical properties exploits the availability of deep multi-band photometry in the VANDELS area, and the measurement of emission lines in the UV rest-frame range (Ly$\alpha$, \ciii, \heii, \oiii). The spectro-photometric approach adopted here is the same used for galaxies in the EoR \citep[e.g.,][]{Stark2015b,Castellano2022a} and provides a useful reference to infer \xion~from observed properties of very high-redshift galaxies.
We explored the correlation between ionizing efficiency and galaxy properties evaluating their statistical significance on the basis of \citet[][]{Spearman04} rank coefficients which are summarised in Fig.~\ref{fig_spearman} both for the entire sample and the high-mass one.

We find no evolution of \xion~with redshift within the probed range, and a mild increase of the ionizing efficiency at fainter $M_{UV}$, and bluer UV slopes.
The most significant correlations are found with respect to EW(Ly$\alpha$), stellar mass, $\Sigma_{SFR}$, and specific SFR. The latter relation is particularly interesting: it is apparently unimodal with a remarkably low scatter, and it is significant both in the full sample and in the log(M$_{star}$/M$_{\odot}$)$>$9.5 one. As a result, the objects above the main-sequence of star-formation consistently have higher-than-average \xion, and, viceversa low \xion~is found in galaxies that have lower SFR than the typical value at their stellar mass. 
Our results can be clearly visualized by looking at the differences between the distributions for objects with high (\logxion$>$25.2) and low efficiency in our high-mass subsample (Fig.~\ref{fig_ionizers}). The probability density distributions of sSFR, SFR and $\Sigma_{SFR}$ are clearly different. A difference is also apparent in the high-EW tail of the EW(Ly$\alpha$) distribution while $M_{UV}$ and UV slope distributions are very similar for the two samples. In fact, according to a two-sided Kolmogorov-Smirnov test \citep[][]{Hodges1958}, the null hypothesis that the two samples are drawn from the same parent distribution has $p<0.01$ for sSFR, SFR and $\Sigma_{SFR}$, and just a slightly higher significance ($p\sim$0.03) in the EW(Ly$\alpha$) case. The subsamples of objects with low and high \xion are statistically equivalent ($p>0.15$) as far as $M_{UV}$ and $\beta$ are concerned. An inverse correlation between metallicity and \xion provides a possible physical explanation for the observed relations \citep{Yung2020}. In such a scenario, the inverse trend of \xion with stellar mass is a natural consequence of the underlying mass-metallicity relation \citep[e.g.,][]{Calabro2021,Curti2023}, while objects with high $\Sigma_{SFR}$ and galaxies above the main sequence have enhanced ionising efficiency due to ongoing star-formation episodes from low-metallicity gas \citep[]{Amorin2017}. The increase in the EW of \ciii~and \oiii~at low gas-phase metallicity \citep[e.g.,][]{Stark2014} also explains the correlations we found between these collisionally excited emission lines and \xion. In turn, the lack of correlation with \heii~is consistent with the findings by \citet{Saxena2020} that high-redshift HeII-emitters and non-emitters have comparable metallicity, and additional, poorly known mechanisms such as X-ray binaries or stripped stars are needed to fully account for the emission rate of HeII ionising photons.

These findings have important consequences for the investigation of the epoch of reionization. 
There are intriguing similarities between our results and observed trends between escape fraction and galaxy properties. The escape fraction has been found to be positively correlated with EW(Ly$\alpha$), sSFR and $\Sigma_{SFR}$, and to be anti-correlated with stellar mass and metallicity \citep[e.g.,][]{Pahl2021,Flury2022,Begley2022}. While differences remain (e.g., the anti-correlation with UV slope being more evident for $f_{esc}$ than for \xion), the emerging scenario points to compact, high sSFR galaxies being efficient sources of ionizing photons that are leaked into the IGM through density-bounded regions or through channels carved by star-formation feedback \citep{Gazagnes2020}. In this respect, the increase in sSFR which has been found at high-redshift \citep[e.g.,][]{Stark2013,Castellano2017,Topping2022} suggests an increase in both ionizing efficiency and in the availability of moderate escape fractions of ionizing photons to keep the IGM ionized \citep{Chisholm2022,Lin2023,Mascia2023b}. 
Finally, both the spectro-photometric analysis of the UV rest-frame range and the availability of sSFR and $\Sigma_{SFR}$ as proxies for \xion~can be of fundamental importance to determine the role at the onset of reionization of the galaxy populations that are being discovered by JWST NIRCam at z$\gtrsim$10 where rest-frame optical emission lines fall outside the spectral range observable with NIRSpec \citep[e.g.,][]{Castellano2022b,Castellano2023a,Naidu2022b,Harikane2022b,Curtis-Lake2022}. The fitting equation presented in Sect.~\ref{subsec:fitproperties} to estimate \xion from SFR and M$_{star}$ provides a first step in this direction. Forthcoming JWST spectroscopic surveys will allow us to extend the present analysis to higher redshifts and lower masses in order to derive robust estimators of the ionizing efficiency of the very first galaxies.

\begin{acknowledgements}
We thank the referee for the detailed and constructive comments. We thank Irene Shivaei for kindly providing tabulated data from \citet[][]{Shivaei2018}.
The present paper exploits Cineca computing resources obtained under projects INA20\_C6T27, and INA20\_C7B36. We acknowledge the computing centre  of Cineca and INAF, under the coordination of the "Accordo Quadro MoU per lo svolgimento di attività congiunta di ricerca Nuove frontiere in Astrofisica: HPC e Data Exploration di nuova generazione", for the availability of computing resources and support. MC and PS acknowledge support from INAF Minigrant ``Reionization and fundamental cosmology with high-redshift galaxies".
\end{acknowledgements}

\bibliographystyle{aa}

\appendix
\section{Average ionizing efficiency as a function of observed and physical properties}
We report in Tab.~\ref{tab:obsproperties} and \ref{tab:physproperties} the average and standard deviation of the ionizing efficiency of the VANDELS objects with log(M$_{star}/M_{\odot})>9.5$ as a function of observed and physical properties. The statistics as a function of mass on the full sample is included in Tab.~\ref{tab:physproperties} for reference.
\\
\begin{table}[ht]
\caption{Ionizing efficiency as a function of observed properties}\label{tab:obsproperties}
\centering
\renewcommand{\arraystretch}{1.3}
\begin{tabular}{ l l l l}
\hline
Min & Max & Median & log($\xi_{ion}^*$) \\
& & & [Hz erg$^{-1}$]\\
\hline
\multicolumn{4}{c}{Redshift}\\
\hline
  2.0 &   2.5 &  2.39 &  25.04$^{+0.26}_{-0.72}$ \\
  2.5 &   3.0 &  2.63 &  24.96$^{+0.26}_{-0.78}$ \\
  3.0 &   3.5 &  3.25 &  24.93$^{+0.25}_{-0.64}$ \\
  3.5 &   4.0 &  3.70 &  24.89$^{+0.24}_{-0.59}$ \\
  4.0 &   4.5 &  4.20 &  25.04$^{+0.24}_{-0.57}$ \\
  4.5 &   5.0 &  4.72 &  24.98$^{+0.20}_{-0.39}$ \\
\hline
\multicolumn{4}{c}{$M_{UV}$}\\
\hline
 -22.0 & -21.0 & -21.29 &  24.94$^{+0.27}_{-0.80}$ \\
 -21.0 & -20.0 & -20.59 &  24.99$^{+0.25}_{-0.62}$ \\
 -20.0 & -19.0 & -19.95 &  24.99$^{+0.20}_{-0.39}$ \\
\hline
\multicolumn{4}{c}{$\beta$}\\
\hline
  -3.0 &  -2.5 & -2.65 &  24.88$^{+0.16}_{-0.25}$ \\
  -2.5 &  -2.0 & -2.10 &  24.97$^{+0.22}_{-0.45}$ \\
  -2.0 &  -1.5 & -1.67 &  25.02$^{+0.24}_{-0.61}$ \\
  -1.5 &  -1.0 & -1.32 &  24.93$^{+0.26}_{-0.78}$ \\
  -1.0 &  -0.5 & -0.78 &  24.92$^{+0.24}_{-0.59}$ \\
\hline
\multicolumn{4}{c}{EW(Ly$\alpha$) (\AA)}\\
\hline
 -40.0 &    0.0 &  -5.83 &  24.96$^{+0.26}_{-0.70}$ \\
 0.0 &   40.0 &  10.47 &  24.96$^{+0.24}_{-0.58}$ \\
 40.0 &   80.0 &  56.41 &  25.06$^{+0.18}_{-0.31}$ \\
 80.0 &  120.0 &  98.36 &  25.34$^{+0.20}_{-0.37}$ \\
\hline
\end{tabular}
\end{table}

\begin{table}[t]
\caption{Ionizing efficiency as a function of physical properties}\label{tab:physproperties}
\centering
\renewcommand{\arraystretch}{1.3}
\begin{tabular}{ l l l l }
\hline
Min & Max & Median & log($\xi_{ion}^*$) \\
& & & [Hz erg$^{-1}$]\\
\hline
\multicolumn{4}{c}{log(M$_{star}/M_{\odot}$)$^a$}\\
\hline
  7.5 &   8.5 &   8.38 &  25.40$^{+0.13}_{-0.19}$ \\
  8.5 &   9.5 &   9.19 &  25.21$^{+0.20}_{-0.37}$ \\
  9.5 &  10.5 &   9.88 &  24.96$^{+0.25}_{-0.64}$ \\
 10.5 &  11.5 &  10.63 &  24.86$^{+0.27}_{-0.92}$ \\
\hline
\multicolumn{4}{c}{log(SFR) [$M_{\odot}/yr$]}\\
\hline
  0.0 &   0.5 &  0.38 &  24.79$^{+0.09}_{-0.12}$ \\
  0.5 &   1.0 &  0.86 &  24.75$^{+0.19}_{-0.35}$ \\
  1.0 &   1.5 &  1.29 &  24.88$^{+0.21}_{-0.43}$ \\
  1.5 &   2.0 &  1.69 &  25.04$^{+0.24}_{-0.60}$ \\
  2.0 &   2.5 &  2.18 &  25.27$^{+0.18}_{-0.31}$ \\
  2.5 &   3.0 &  2.60 &  25.42$^{+0.07}_{-0.09}$ \\
\hline
\multicolumn{4}{c}{log(sSFR) [yr$^{-1}$]}\\
\hline
-10.0 &  -9.5 & -9.63 &  24.57$^{+0.24}_{-0.61}$ \\
 -9.5 &  -9.0 & -9.09 &  24.70$^{+0.25}_{-0.65}$ \\
 -9.0 &  -8.5 & -8.71 &  24.78$^{+0.19}_{-0.35}$ \\
 -8.5 &  -8.0 & -8.34 &  25.05$^{+0.20}_{-0.37}$ \\
 -8.0 &  -7.5 & -7.90 &  25.35$^{+0.10}_{-0.12}$ \\
 -7.5 &  -7.0 & -7.44 &  25.45$^{+0.08}_{-0.10}$ \\
\hline
\multicolumn{4}{c}{log($\Sigma_{SFR}$) [$M_{\odot}/yr/kpc^2$]}\\
\hline 
  -1.0 &  -0.5 & -0.68 &  24.72$^{+0.17}_{-0.28}$ \\
  -0.5 &   0.0 & -0.18 &  24.67$^{+0.15}_{-0.24}$ \\
   0.0 &   0.5 &  0.28 &  24.93$^{+0.22}_{-0.47}$ \\
   0.5 &   1.0 &  0.75 &  25.01$^{+0.23}_{-0.51}$ \\
   1.0 &   1.5 &  1.20 &  25.11$^{+0.22}_{-0.46}$ \\
   1.5 &   2.0 &  1.66 &  25.33$^{+0.18}_{-0.33}$ \\
\hline
\end{tabular}
\small \\a) The VANDELS sample is incomplete at log(M$_{star}/M_{\odot})<9.5$ (Sec.~\ref{sec:BEAGLE}), the average and standard deviation in this mass range are included here for reference.
\end{table}

\end{document}